\documentclass[aip,jap,amsmath,amssymb,reprint]{revtex4-1}

\usepackage{graphicx}
\usepackage{dcolumn}
\usepackage{bm}
\usepackage{amsfonts}
\usepackage{amsmath}
\usepackage{amssymb}
\usepackage{subfigure}
\usepackage{CJK}
\usepackage{mathptmx}
\usepackage[breaklinks,colorlinks,linkcolor=blue,citecolor=blue,urlcolor=black,bookmarks=false]{hyperref}
\usepackage{placeins}

\def\unit#1{\mathord{\thinspace\rm #1}}

\begin{document}
	
	
	\title{Few-layer graphene patterned bottom gates for van der Waals heterostructures}%

\author{Martin Drienovsky}%
\email{martin.drienovsky@physik.uni-regensburg.de}
\affiliation{Institute of Experimental and Applied Physics, University of Regensburg, D-93040 Regensburg, Germany}%
\author{Andreas Sandner}
\affiliation{Institute of Experimental and Applied Physics, University of Regensburg, D-93040 Regensburg, Germany}%
\author{Christian Baumgartner}
\affiliation{Institute of Experimental and Applied Physics, University of Regensburg, D-93040 Regensburg, Germany}%
\author{Ming-Hao Liu}
\altaffiliation{new address: Department of Physics, National Cheng Kung University, Tainan 70101, Taiwan}%
\affiliation{Institute of Theoretical Physics, University of Regensburg, D-93040 Regensburg, Germany}%
\author{Takashi Taniguchi}
\affiliation{National Institute for Materials Science, 1-1 Namiki, Tsukuba 305-0044, Japan}%
\author{Kenji Watanabe}
\affiliation{National Institute for Materials Science, 1-1 Namiki, Tsukuba 305-0044, Japan}%
\author{Klaus Richter}
\affiliation{Institute of Theoretical Physics, University of Regensburg, D-93040 Regensburg, Germany}%
\author{Dieter Weiss}
\affiliation{Institute of Experimental and Applied Physics, University of Regensburg, D-93040 Regensburg, Germany}%
\author{Jonathan Eroms}%
\affiliation{Institute of Experimental and Applied Physics, University of Regensburg, D-93040 Regensburg, Germany}%

\date{\today}

\begin{abstract}
We introduce a  method of local gating for van der Waals heterostructures, employing a few-layer graphene patterned bottom gate. Being a member of the 2D material family, few-layer graphene adapts perfectly to the commonly used stacking method. Its  versatility regarding patterning as well as its flatness make it an ideal candidate for experiments on locally gated 2D materials. Moreover, in combination with ultra-thin hexagonal boron nitride as an insulating layer, sharp potential steps can be created and the quality of the investigated 2D material can be sustained. To underline the good feasibility and performance, we show results on transport experiments in periodically modulated graphene- boron nitride heterostructures, where the charge carrier density is tuned via locally acting patterned few layer graphene bottom gates and a global back gate. 
\end{abstract}

\maketitle

\section{Introduction}
\label{sec:1}
Since its discovery a decade ago \cite{novoselov2004,novoselov2005,geim2007progress}, graphene has been spearheading a steadily growing zoo of new 2D materials\cite{xu2013_2Dmaterials}. Heterostructures of these atomically thin crystals open up an exciting playground for tailoring unique material properties\cite{dean2010hbntransfer, geim2013van, novoselov2016_2d_vdW}. In order to explore the energy landscapes of these stacks, one can rely on electrostatic gating that allows to move through their energy spectrum. Here, planar conductors covered with an insulator of variable thickness provide global control of the Fermi-level of the material placed on top and thus enable the tuning of the charge carrier density. Additionally, the introduction of local gating permits to investigate the effects of varying potential on the micro- or even nanoscale. \\
In this work, we focus on graphene, where a local potential modulation can be imposed either by chemical gating\cite{baringhaus2015ChemicalGating}, geometrical variation \cite{han2007gnr, Tombros2011} local perforation \cite{sandner2015ballistic, Yagi2015} or gate electrodes, the latter method being the most versatile one regarding tunability.  Local gating in graphene can generate pn-junctions via the electric field effect and yield unique effects such as Klein tunneling \cite{klein1929reflexion, katsnelson2006chiral}, Klein collimation\cite{cheianov2006selective} and lensing\cite{cheianov2007focusing, chen2016electronOptics, liu2017lensing}. Yet, tunable pn-junctions are not only restricted to graphene, as in other atomically thin, ambipolar materials like black phosphorus or WS$_2$, bipolar transitions can also be locally induced \cite{buscema2014photovoltaic,baugher2014WSe2_pn-diode}. 
In early studies on charge carrier density-modulated graphene, metallic top gates, separated from the graphene\cite{young2009quantum, stander2009evidence, williams2007quantum, ozyilmaz2007electronic, drienovsky2014multibarriers} by a deposited dielectric such as alumina were a common tool of choice. However, electron mobility remained limited due to roughness of the substrate, resist residues and impurities\cite{fallahazad2010impurities}. Subsequently, graphene encapsulation between hexagonal boron nitride (hBN) layers\cite{wang2013oneDcontacts} emerged and drastically increased the common sample quality.

Fabrication of metal top gates on graphene-boron nitride heterostructures however can be a challenge. Usually, the samples require a further side-passivation step to prevent shortcuts between the bare graphene edges and the top gate. Even though structures on the micrometer scale remain stable, narrow metal stripes of only a few tens of nanometers width tend to move on the chemically inert hBN-surface, or they rip when evaporated over the edges of the mesa. This drastically decreases sample yield, since metalization is the last critical fabrication step here. A good alternative, not only for transport, but also for optical experiments, are local bottom gates. In contrast to top gates, local bottom gates do not block or modulate incoming light waves. Among various geometries and choice of metal, planar or step-like graphite or few layer graphene bottom gates have also been reported in different publications\cite{ponomarenko2011DoubleLayers,hunt2013massiveDF_Hofstadter, chen2016electronOptics}. If not embedded\cite{nam2011embedded_gates} into the insulating substrate, metal structures or graphite have a certain height and thin 2D materials transfered on top may bend at the edges which can yield strain\cite{pereira2009strain}. In contrast, suspended graphene is not affected by step-like gates and above all shows excellent quality\cite{rickhaus2013ballistic,grushina2013ballistic,oksanen2013single,rickhaus2015guiding}. However it is constrained to very smooth potential variations, since the gate-to-graphene distance is relatively large and a short periodic modulation is out of reach.\\
Here, we present a new versatile method of bottom gating, employing patterned few layer graphene (FLG). Our method relies on the atomic flatness and stability of FLG and hence minimizes bending or strain of the investigated 2D heterostructure on top. It further enables sharp potential steps and via patterning, one can create any tunable 1D or 2D periodic potential landscape on the nanoscale.

\section{Sample fabrication}
\label{sec:3}

For the preparation of few layer graphene patterned bottom gates (PBG) we select FLG flakes via optical microscopy. In contrast to evaporated metal layers, the thickness of FLG is on the order of a nanometer, and the material is still well conducting. So the minimal height and the eminent flatness of the FLG allow us to neglect any spatial perturbation of the heterostructure transferred on top of the PBG.  For the structures presented within this work, we employ 2-3-layer graphene to assure full screening of the global back gate electric field, formed by the doped Si substrate. The FLG is subjected to electron beam lithography (EBL) and subsequently gets etched into the required shape and pattern by oxygen plasma reactive ion etching (RIE). We select clean and regular gate structures already before the transfer of the heterostructure. Above all, the FLG can be shaped into any desired structure - from periodic, lateral stripes to a parabolic gate electrode that can serve as an electron focusing lens\cite{liu2017lensing}. 

\begin{figure}[t]
	\includegraphics[width=0.95\columnwidth]{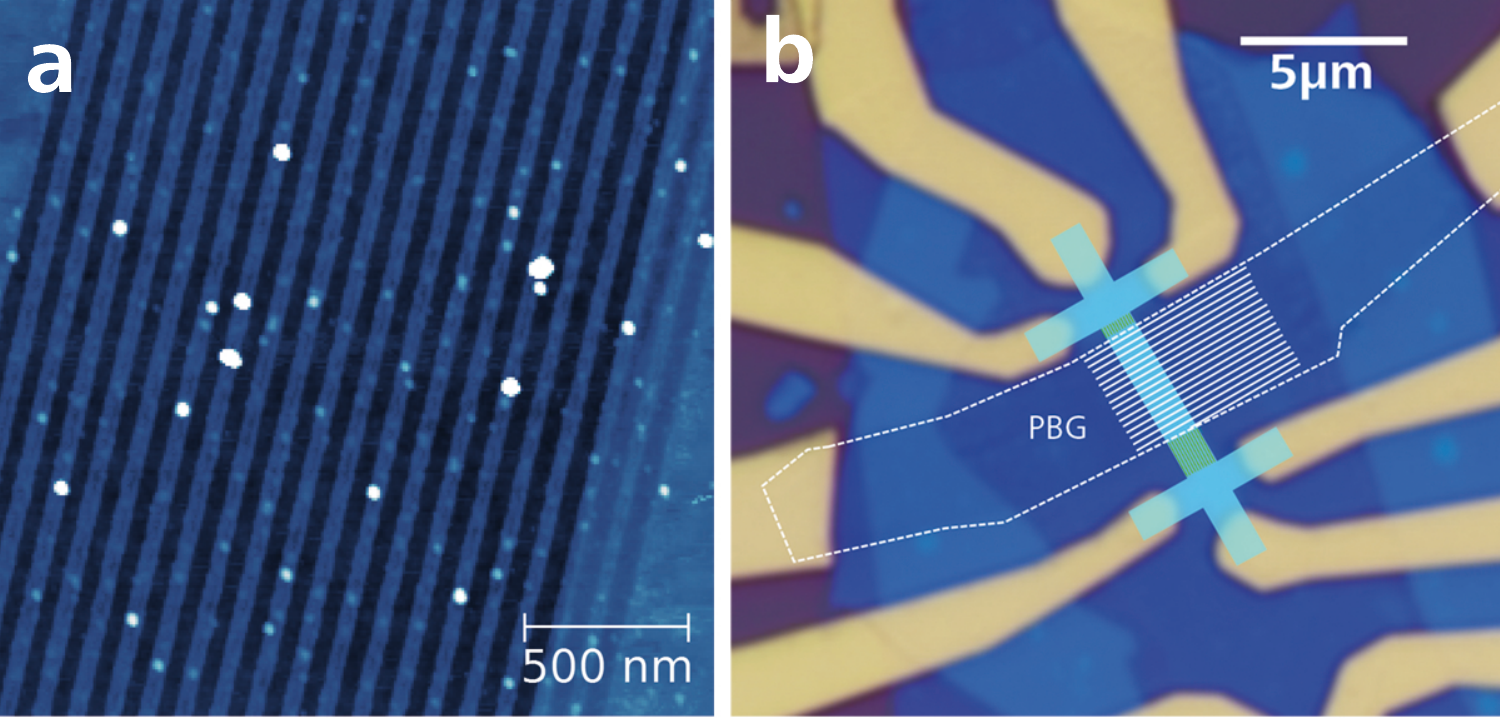}
	\caption{\textbf{Few layer graphene patterned bottom gate and contacted heterostructure.} \textbf{a)} AFM image of a trilayer-graphene (3LG) PBG on a SiO$_2$ substrate. The 3LG has been patterned by oxygen plasma RIE. Resist residues appear in white. The period of the shown PBG is $a=100\unit{nm}$ and the stripe width is $w=50\unit{nm}$. The height of the gate stripes is $\approx 2\unit{nm}$. \textbf{b)} An optical microscope image of a 1D periodic modulated sample, employing a PBG, consisting of 19 FLG stripes beneath the Hall bar. The colored inserts of the Hall bar (cyan) and PBG (white) serve as a guide to the eye. 
	}
	\label{pic:fig1}
\end{figure}

\begin{figure*}[t]
	\includegraphics[width=1.95\columnwidth]{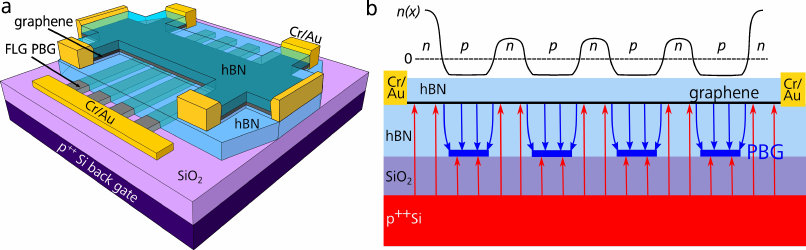}
	\caption{\textbf{Schematic of the sample geometry and a longitudinal cross section.} \textbf{a)} Graphene is encapsulated between two thin hBN sheets and etched into a Hall bar shape. The etching requires depth control so that the PBG is not harmed. The graphene is side-contacted by Cr/Au. The lower hBN adapts to the weak height modulation of the PBG, which is buried below and also contacted with Cr/Au. The global back gate is separated from the PBG by SiO$_2$. \textbf{b)} The cross section shows the different layers and the field lines of gate electrodes. In this example, the field from the positive global gate is screened by the PBG stripes. The locally acting PBG is negatively charged and together the gates generate a bipolar multibarrier. The schematic of the expected longitudinal density profile $n(x)$ is shown above.  
	}
	\label{pic:fig2}
\end{figure*}

An example of a finished 3-layer graphene PBG is depicted in Fig. \ref{pic:fig1}\textbf{a}, consisting of a $100\unit{nm}$-periodic lattice with a stripe width of $w=50\unit{nm}$. The average profile height is $\approx 2\unit{nm}$ \footnote{which would be six times the interlayer constant of graphene. This discrepancy can either be attributed to an overestimation of the height by the atomic force microscope (AFM) or a physical etching of the SiO$_2$ surface by O$_2$-plasma at higher powers and durations}. For the heterostructure we adapted the dry transfer method, following the work of Dean \textit{et al.}\cite{dean2010hbntransfer} and Wang \textit{et al.}\cite{wang2013oneDcontacts}. After a careful choice of hBN and graphene flakes, we stack them on top of each other in a geometry that allows us to contact the conducting layers of the van der Waals heterostructure and the PBG independently. Employing  EBL and RIE (CHF$_3$+O$_2$), the stack can be shaped into a Hall bar, or any other gate geometry of interest and edge-contacted with Cr/Au leads. Note that the precise control of the etch depth is critical especially for rather thin bottom hBN flakes in order not to damage the PBG layer (see Fig. \ref{pic:fig2})\\

\section{Experiments on patterned FLG bottom gates}
\label{sec:4}

\subsection{1D periodic lateral density modulation}

To demonstrate the feasibility of the new gating scheme, we first prepare a dense array of narrow FLG stripes (c.f. Fig. \ref{pic:fig1}\textbf{a}) and create a periodically, density modulated graphene multibarrier system, which has been investigated in samples with much lower mobility in earlier works\cite{dubey2013tunable, drienovsky2014multibarriers}. Figure \ref{pic:fig2}\textbf{a} depicts the general geometry, where the lateral modulation of charge carrier density can be achieved via the interplay of the planar, global back gate and the few layer graphene PBG. The PBG tunes the charge carrier density in the areas above the gate stripes and effectively screens the back gate. Between the stripes, the electric fields of the PBG and the global Si back gate superpose. Outside the PBG-region, the density is solely controlled by the Si back gate (Fig. \ref{pic:fig2}\textbf{b}). Employing ultra-thin hBN as the insulating layer between the graphene and PBG, the effect of stray fields can be reduced and a sharp potential profile achieved. 

\begin{figure}[t]
	\includegraphics[width=0.95\columnwidth]{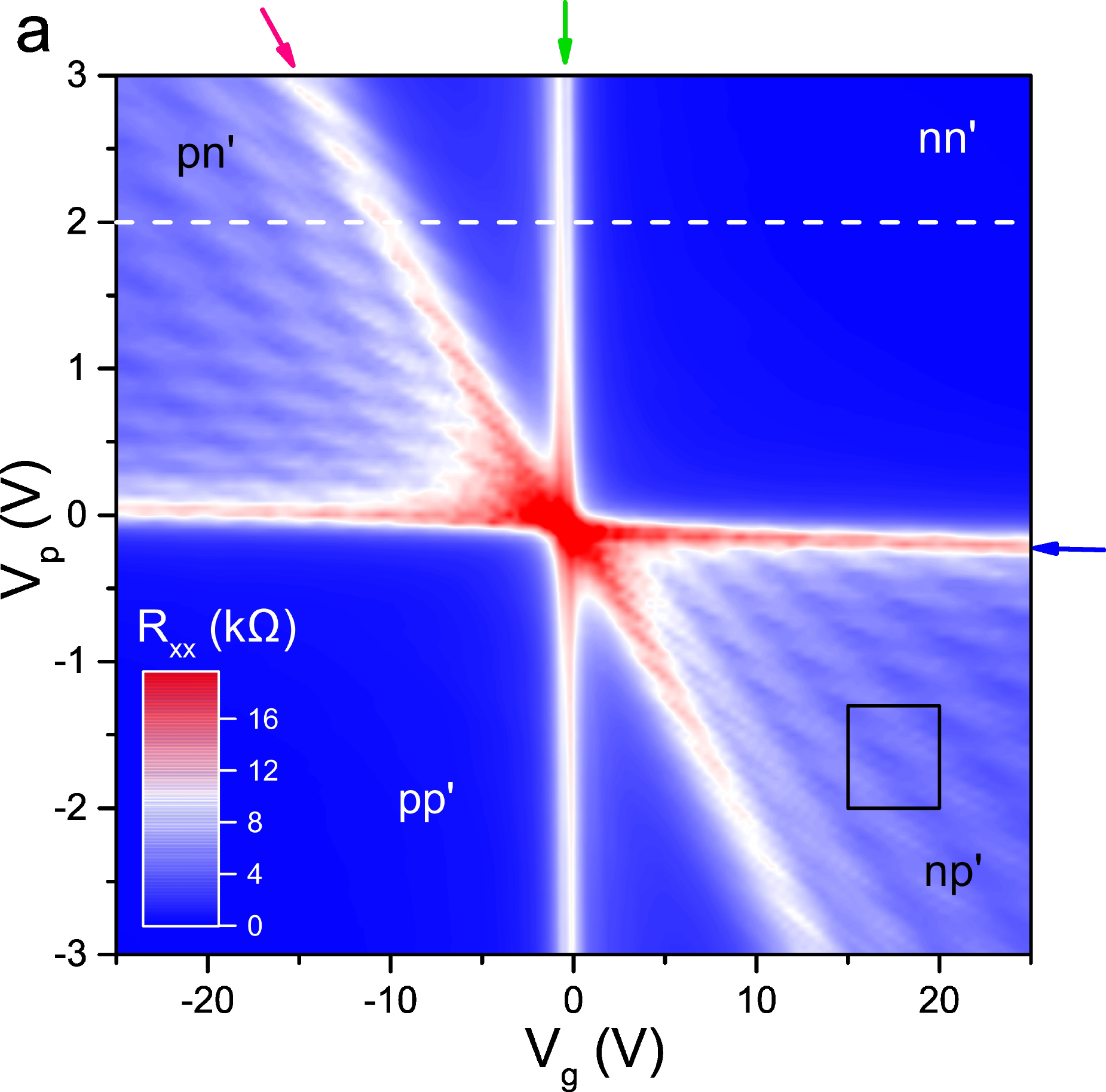}
	\includegraphics[width=0.52\columnwidth]{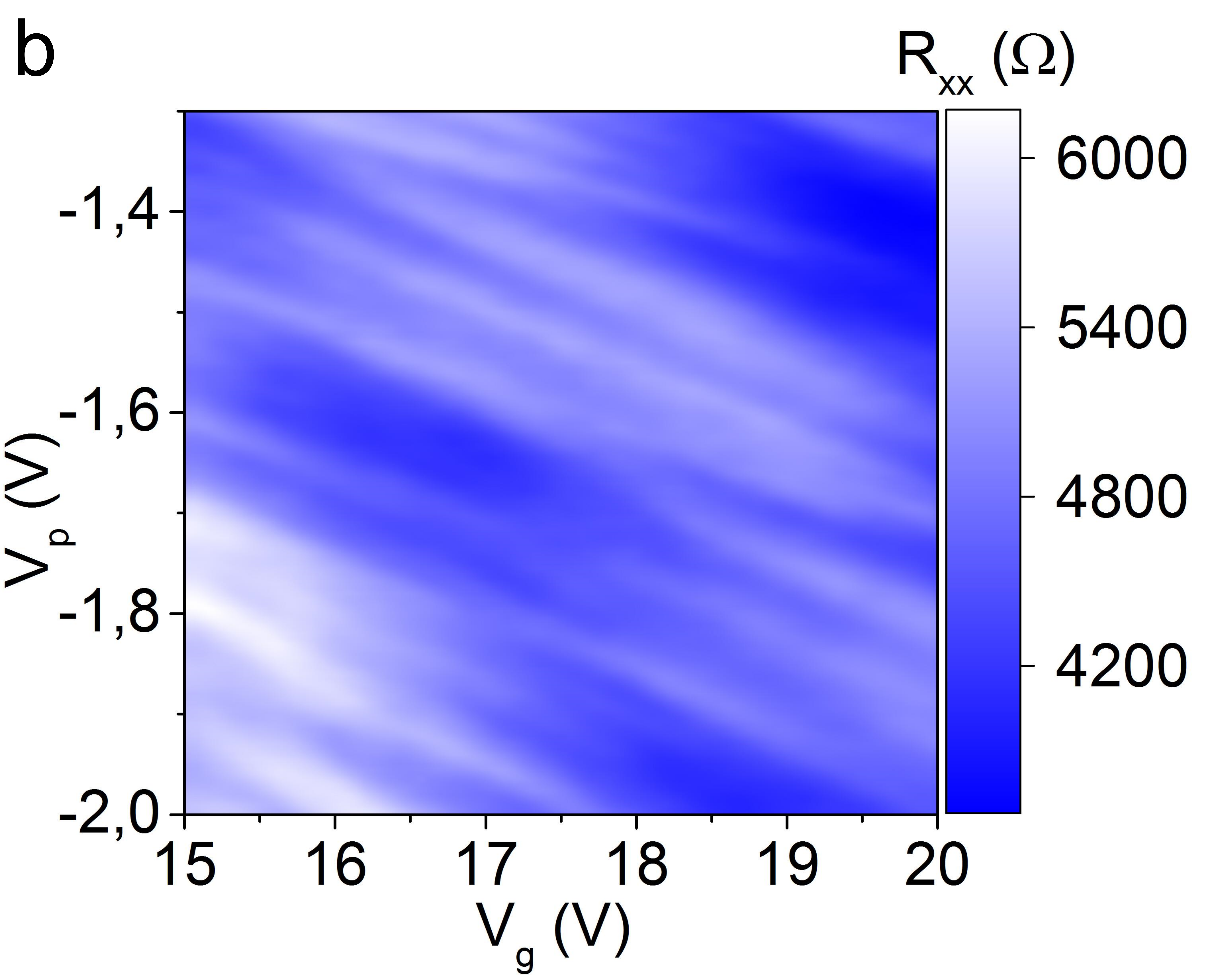}
	\includegraphics[width=0.45\columnwidth]{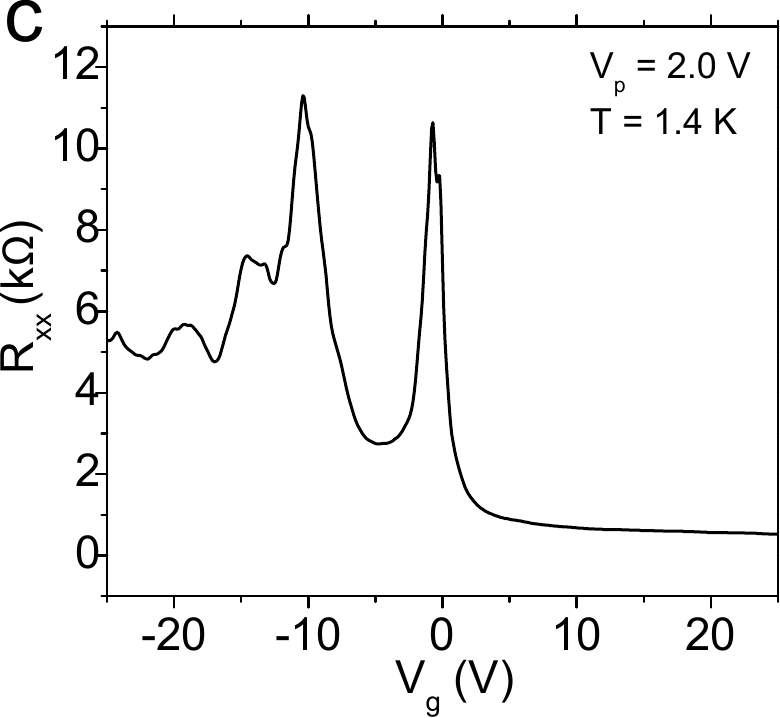}
	\caption{\textbf{Experimental data of a 1D laterally density modulated sample}.\textbf{a)} In the color coded resistance map one can distinguish 4 quadrants, $pn^\prime$, $nn^\prime$, $pp^\prime$, $np^\prime$, where $n$ is the charge carrier density in the (globally gated) area between the gate stripes and $n^\prime$ stands for the locally gated PBG region. In the bipolar regime, a clear Fabry-P\'erot interference pattern can be observed. \textbf{b)} Furthermore, we observe an oscillation of higher frequency in the zoom-in (black box in \textbf{a}) within the basic FP mesh. \textbf{c)} R$_{xx}$, following the white dashed line cut from the resistance map \textbf{a}. 
	}
	\label{pic:fig3}
\end{figure}

In the following, we show experimental data and simulations on encapsulated graphene with a periodic stripe gate of $a=200\unit{nm}$ and $w=100\unit{nm}$. We apply a current of $10\unit{nA}$ and measure the longitudinal (4-point) resistance $R_{xx}$ in a He-4 cryostat at a temperature of $1.4\unit{K}$ within a standard Lock-In setup. The electron mobility $\mu$, derived from the Dirac peak of the outer sample area  (c.f. Dirac peak in Fig. \ref{pic:fig3}\textbf{c} at $V_{g}\approx 0$) is estimated to be $\sim 130 000\unit{cm^2/Vs}$. Here, only the classical gate capacitance $C=\epsilon\epsilon_0 A/d$ of the SiO$_2$ ($d=285\unit{nm}$) and hBN ($d=13\unit{nm}$) has been considered. An estimate of $\unit\mu$ in the PBG area, using the carrier density derived from the Shubnikov-de Haas oscillations in $R_{xx}$ at high magnetic field gives $\sim 45 000\unit{cm^2/Vs}$ and hence a mean free path of $l_m \approx 800\unit{nm}$ at $V_g = 25\unit{V}$. The resulting $R_{xx}(B=0)$ map with respect to the voltages of patterned gate $V_{p}$ and the global back gate $V_{g}$ is displayed in Fig. \ref{pic:fig3}\textbf{a}. We obtain 4 quadrants of quite different resistance ($pn^\prime$, $nn^\prime$, $pp^\prime$, $np^\prime$) that originate from the different charge carrier density and/or polarity of the differently gated areas. The vertical maximum in Fig.~\ref{pic:fig3}\textbf{a} (indicated by a green arrow) is the charge neutrality (Dirac) line of the outer regions (c.f. Fig.\ \ref{pic:fig2}\textbf{b}) that are controlled by the Si back gate. The horizontal line (blue arrow) reflects the charge neutrality point on the areas above the PBG stripes. The vanishing slope of this line indicates that the PBG almost perfectly screens the back gate field. This can be an advantage over top gated samples, where the back gate always shifts the energy homogeneously over the whole sample area\cite{drienovsky2014multibarriers}. The diagonal Dirac line (magenta arrow) originates from the area between the PBG stripes, where charge carrier density is primarily controlled by the back gate, but is also PBG-affected due to stray fields at the stripe edges (Fig. \ref{pic:fig2}\textit{b}). 

In the bipolar transport regime, gate-tunable cavities emerge due to angle dependent Klein tunneling (top of Fig. \ref{pic:fig2}\textbf{b}). Consequently, we encounter two sets of Fabry-P\'erot (FP)-resonances (parallel to the diagonal and horizontal Dirac lines, respectively) that interfere and give rise to a rhombic resistance pattern. From its regularity we can conclude that the cavities are of identical shape and width. A simple resonance condition\cite{drienovsky2014multibarriers} ignoring Klein tunneling gives already an acceptable agreement of the basic FP-oscillation frequency with the real spatial properties of the PBG. We extract a cavity width of roughly $100\unit{nm}$, which corresponds well to the PBG-stripe width of $\approx100\unit{nm}$. In addition to the well known bipolar FP-oscillations, we observe a fine resonance pattern within the rhombic mesh (Fig.\ \ref{pic:fig3}\textbf{b}) that does not correspond to any cavity in the superlattice. However, this new set of oscillations is reproducible in different samples with similar superlattice parameters and also becomes visible in a quantum transport simulation that will be discussed in more detail below. Since we find good consistency between experiment and theory, we attribute the emergence of these fine oscillations to ballistic transport across several potential barriers. 


We compare the experimental results to a quantum transport simulation based on a real-space tight-binding model with periodic boundary condition along the lateral dimension applied \cite{liu2012quantum_simulation} and a realistic on-site energy profile $V(x)$. $V(x)$ is extracted from the gate-induced carrier density, simulated by a finite-element-based electrostatic simulation, employing the exact gating scheme of the two interacting gates and including quantum capacitance, similar to Ref.~\cite{drienovsky2014multibarriers}. The clean part of the Hamiltonian has been further scaled by a factor of 4 following Ref.~\cite{liu2015tight_binding} for speeding up the calculations. Due to the finite phase-coherence length of the experiment, we consider a $1.6$-$\unit{\mu m}$-long effective two-terminal model covering only 4 periods of the PBG with the Fermi energy in the incoming and outgoing leads fixed at $0.1\unit{eV}$. Contrary to Refs.~\cite{liu2012quantum_simulation, rickhaus2013ballistic}, we only compute the normalized conductance $g$ without performing the mode counting and concentrate on a qualitative comparison of the interference pattern.

\begin{figure}
	\includegraphics[width=0.95\columnwidth]{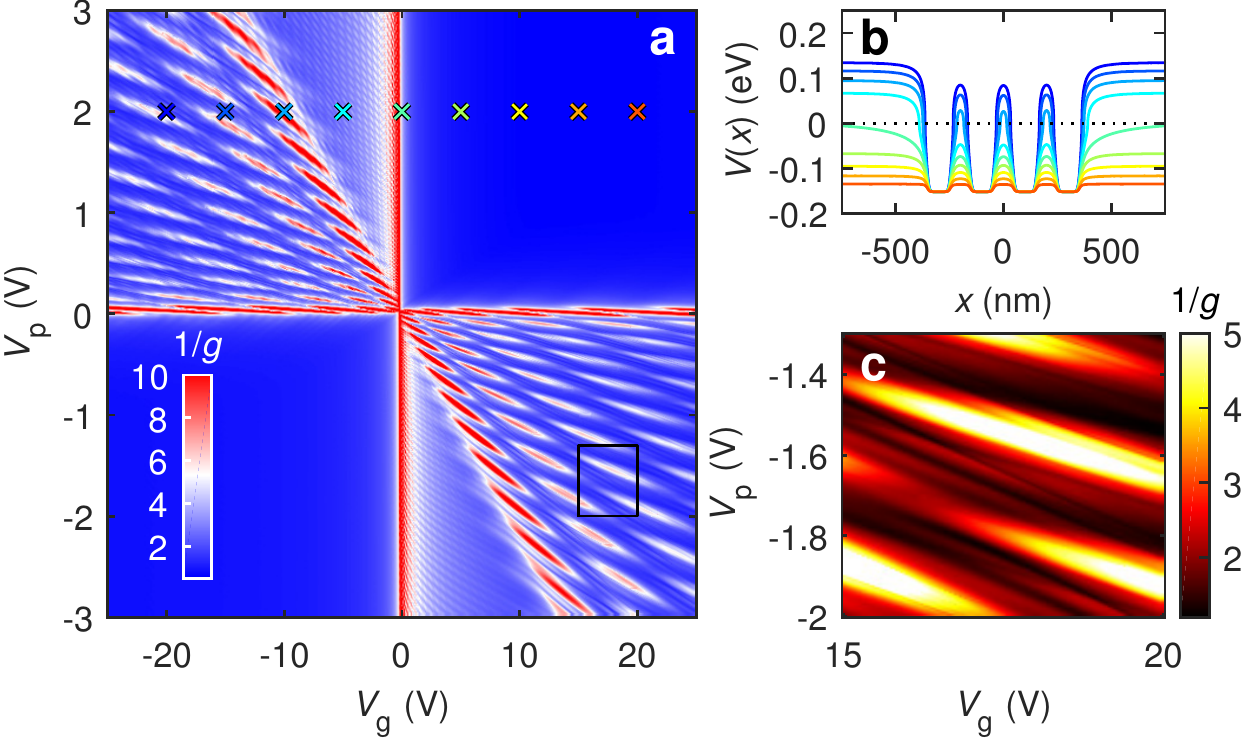}
	\caption{\textbf{Quantum transport simulation for the 1D modulated sample.} \textbf{a)} Inverse of the normalized conductance $1/g$ as a function of $V_g$ and $V_p$, based on an effective model with four periods of the PBG. Examples of the on-site energy profile $V(x)$ are shown in \textbf{b} with their corresponding gate voltages marked by the colored crosses in \textbf{a}. \textbf{c)} Zoom-in of $1/g$ for the region marked by the black box in \textbf{a}.}
	\label{pic:fig5}
\end{figure}

The inverse of the normalized conductance $1/g$ as a function of $V_g$ and $V_p$ is presented in Fig.~\ref{pic:fig5}\textbf{a}, where a few crosses mark the corresponding gate voltages considered in the exemplary profiles of $V(x)$ shown in Fig.~\ref{pic:fig5}\textbf{b}.
A qualitative agreement of the experiment (Fig.~3\textbf{a}) and simulation (Fig.~\ref{pic:fig5}\textbf{a}) is obvious, particularly the position of the charge neutrality lines and the complex FP pattern. 
The coarse FP pattern can be recreated independently of the number of stripes included in the model, since the oscillations mainly depend on resonances within individual cavities. The model calculation is fully ballistic, so in order to recreate the finite scattering length in the experiment, we calculated transmission for a varying number of PBG stripes.
The fine oscillations can be reproduced best by employing 4 PBG stripes (Fig.~\ref{pic:fig5}\textbf{c}), whereas for 3 or 5 stripes the oscillation frequency does not fit to the experiment. We would like to emphasize that the fine oscillations appear in the simulation even if the outer regions of the sample or the leads are not taken into account. Hence resonances from larger cavities can be excluded and we can speak of a multibarrier effect. Since the experiment is best reproduced with 4 periods of PBG stripes, it indicates a ballistic length of $800\unit{nm}$, which is in good agreement with the experimentally derived mean free path. The observation of FP oscillations as well as coherent transport over several lattice periods clearly shows that the method is well suited for preparing high quality potential modulations in graphene on the nanoscale.

\subsection{2D periodically modulated graphene}

\begin{figure}[t]
	\includegraphics[width=0.95\columnwidth]{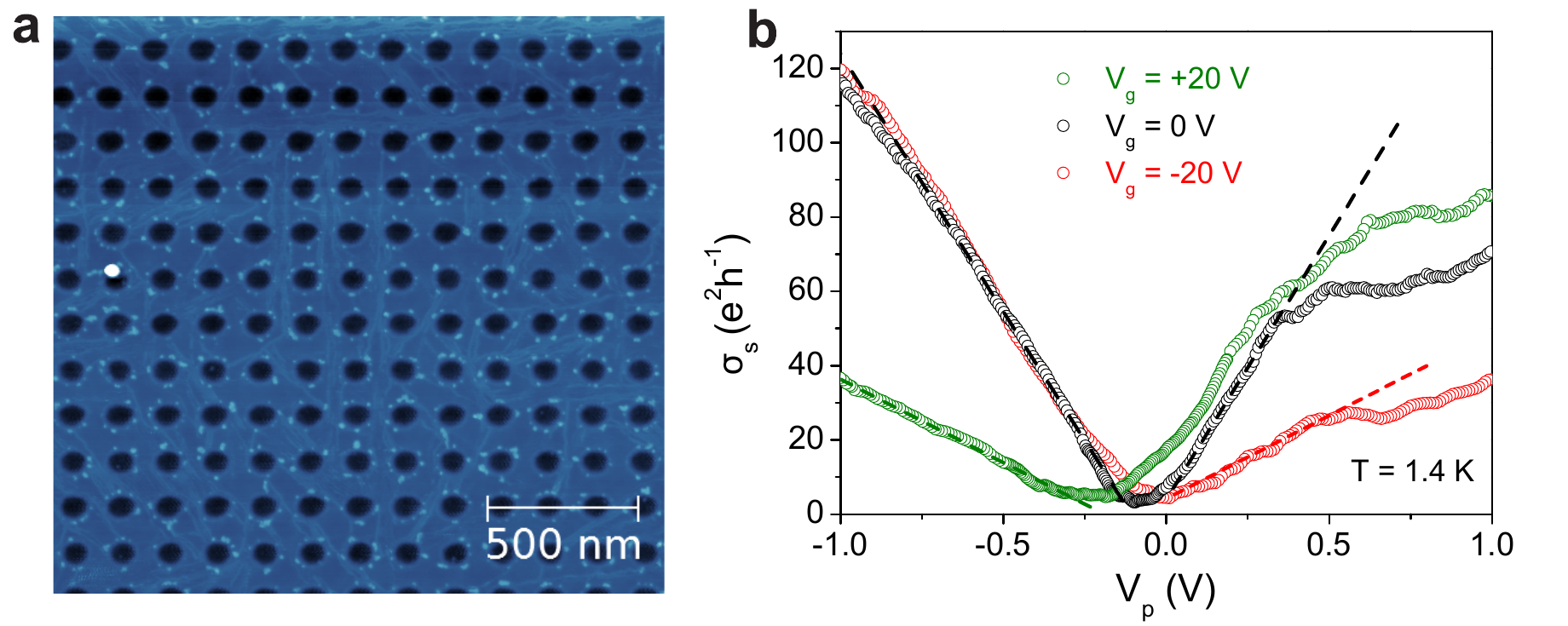}
	\caption{\textbf{2D periodic bottom gate and gate response.} \textbf{a)}, AFM image of a FLG PBG patterned in a  periodic 2D hole-lattice. The period of the array is $a = 150\unit{nm}$ and the diameter is $d = 75\unit{nm}$. \textbf{b)}, 4-point resistance as a function of the  PBG voltage $V_{p}$ at different global gate voltages $V_{g}$ and $T = 1.4\unit{K}$. 
	}
	\label{pic:andi}
\end{figure}

In addition to the 1D laterally patterned gates discussed above, we performed transport experiments on 2D periodically modulated few layer graphene PBG. The fabrication method for these samples was similar, except for the PBG, which now features a 2D periodic array of holes. Figure \ref{pic:andi}\textbf{a} shows an AFM picture of a PBG with a 2D periodic array of etched holes. We successfully measured transport properties of encapsulated graphene on patterned gates with lattice constants of $a = 150 - 300\unit{nm}$ and hole diameters of $d = 50 - 150\unit{nm}$.\\
Again, the charge carrier modulation can be influenced by the interplay between the global back gate and the PBG. In this geometry, the Si back gate affects only the area of the graphene sample above the holes and the rest of the biased graphene sheet is screened by the PBG. The most interesting feature of this sample design is the possibility to tune between the unipolar and bipolar transport regime, i.e. generating an 2D-array of pn-junctions.\\
We probed the 4-point resistance as a function of $V_{p}$ at different global back gate voltages $V_{g}$ (Fig. \ref{pic:andi}b) for a sample with $a = 300\unit{nm}$ and $d = 150\unit{nm}$. The interplay between the two gates has a clear influence on the transport properties of our sample. While the curve for  $V_{g} = 0\unit{V}$ is quite narrow and symmetric, this is no longer true for measurements with applied $V_{g}$. There, we see a difference in the $\sigma(V_p)$-curves  between the unipolar (e.g. $V_g, V_p > 0$) and the bipolar case (e.g. $V_g > 0, V_p < 0$). If we extract the corresponding charge carrier mobilities from the slopes, we get $\mu \approx 10000\unit{cm^2/Vs}$ in the bipolar and $\mu \approx 40000\unit{cm^2/Vs}$ in the unipolar regime, which is essentially the same as in the case of $V_{g}=0\unit{V}$. While the mobility in the bipolar regime is considerably decreased by reflections and scattering at the imposed pn-junctions in the graphene layer, this effect seems to be much weaker in the unipolar case.

\section{New possibilities}
\label{sec:5}
With the few-layer graphene patterned bottom gates, we introduce a new method for locally gating van der Waals heterostructures and 2D materials. The experiments and simulation presented here verify that by employing a few layer graphene PBG, a locally sharp charge carrier density modulation can be achieved as well as a high sample quality sustained. We believe few layer graphene gates can give the opportunity for a variety of new and otherwise impracticable experiments. The great flexibility of the PBG in the fabrication and its easy integration into the van der Waals stacking process allows us to tailor geometries of variable interest and investigate artificial potential landscapes, employing 1D and 2D patterns as well as (curved) bipolar junctions \cite{liu2017lensing}, guides and mirrors. \\

\section*{Acknowledgments}
Financial support by the Deutsche Forschungsgemeinschaft (DFG) within the programs GRK 1570 and SFB 689 is gratefully acknowledged.

\bibliography{GraphenLatexBib}

\begin{thebibliography}{39}%
\makeatletter
\providecommand \@ifxundefined [1]{%
 \@ifx{#1\undefined}
}%
\providecommand \@ifnum [1]{%
 \ifnum #1\expandafter \@firstoftwo
 \else \expandafter \@secondoftwo
 \fi
}%
\providecommand \@ifx [1]{%
 \ifx #1\expandafter \@firstoftwo
 \else \expandafter \@secondoftwo
 \fi
}%
\providecommand \natexlab [1]{#1}%
\providecommand \enquote  [1]{``#1''}%
\providecommand \bibnamefont  [1]{#1}%
\providecommand \bibfnamefont [1]{#1}%
\providecommand \citenamefont [1]{#1}%
\providecommand \href@noop [0]{\@secondoftwo}%
\providecommand \href [0]{\begingroup \@sanitize@url \@href}%
\providecommand \@href[1]{\@@startlink{#1}\@@href}%
\providecommand \@@href[1]{\endgroup#1\@@endlink}%
\providecommand \@sanitize@url [0]{\catcode `\\12\catcode `\$12\catcode
  `\&12\catcode `\#12\catcode `\^12\catcode `\_12\catcode `\%12\relax}%
\providecommand \@@startlink[1]{}%
\providecommand \@@endlink[0]{}%
\providecommand \url  [0]{\begingroup\@sanitize@url \@url }%
\providecommand \@url [1]{\endgroup\@href {#1}{\urlprefix }}%
\providecommand \urlprefix  [0]{URL }%
\providecommand \Eprint [0]{\href }%
\providecommand \doibase [0]{http://dx.doi.org/}%
\providecommand \selectlanguage [0]{\@gobble}%
\providecommand \bibinfo  [0]{\@secondoftwo}%
\providecommand \bibfield  [0]{\@secondoftwo}%
\providecommand \translation [1]{[#1]}%
\providecommand \BibitemOpen [0]{}%
\providecommand \bibitemStop [0]{}%
\providecommand \bibitemNoStop [0]{.\EOS\space}%
\providecommand \EOS [0]{\spacefactor3000\relax}%
\providecommand \BibitemShut  [1]{\csname bibitem#1\endcsname}%
\let\auto@bib@innerbib\@empty
\bibitem [{\citenamefont {Novoselov}\ \emph {et~al.}(2004)\citenamefont
  {Novoselov}, \citenamefont {Geim}, \citenamefont {Morozov}, \citenamefont
  {Jiang}, \citenamefont {Zhang}, \citenamefont {Dubonos}, \citenamefont
  {Grigorieva},\ and\ \citenamefont {Firsov}}]{novoselov2004}%
  \BibitemOpen
  \bibfield  {author} {\bibinfo {author} {\bibfnamefont {K.}~\bibnamefont
  {Novoselov}}, \bibinfo {author} {\bibfnamefont {A.}~\bibnamefont {Geim}},
  \bibinfo {author} {\bibfnamefont {S.}~\bibnamefont {Morozov}}, \bibinfo
  {author} {\bibfnamefont {D.}~\bibnamefont {Jiang}}, \bibinfo {author}
  {\bibfnamefont {Y.}~\bibnamefont {Zhang}}, \bibinfo {author} {\bibfnamefont
  {S.}~\bibnamefont {Dubonos}}, \bibinfo {author} {\bibfnamefont
  {I.}~\bibnamefont {Grigorieva}}, \ and\ \bibinfo {author} {\bibfnamefont
  {A.}~\bibnamefont {Firsov}},\ }\bibfield  {title} {\enquote {\bibinfo {title}
  {Electric field effect in atomically thin carbon films},}\ }\href@noop {}
  {\bibfield  {journal} {\bibinfo  {journal} {Science}\ }\textbf {\bibinfo
  {volume} {306}},\ \bibinfo {pages} {666--669} (\bibinfo {year}
  {2004})}\BibitemShut {NoStop}%
\bibitem [{\citenamefont {Novoselov}\ \emph {et~al.}(2005)\citenamefont
  {Novoselov}, \citenamefont {Geim}, \citenamefont {Morozov}, \citenamefont
  {Jiang}, \citenamefont {Katsnelson}, \citenamefont {Grigorieva},
  \citenamefont {Dubonos},\ and\ \citenamefont {Firsov}}]{novoselov2005}%
  \BibitemOpen
  \bibfield  {author} {\bibinfo {author} {\bibfnamefont {K.~S.}\ \bibnamefont
  {Novoselov}}, \bibinfo {author} {\bibfnamefont {A.~K.}\ \bibnamefont {Geim}},
  \bibinfo {author} {\bibfnamefont {S.~V.}\ \bibnamefont {Morozov}}, \bibinfo
  {author} {\bibfnamefont {D.}~\bibnamefont {Jiang}}, \bibinfo {author}
  {\bibfnamefont {M.~I.}\ \bibnamefont {Katsnelson}}, \bibinfo {author}
  {\bibfnamefont {I.~V.}\ \bibnamefont {Grigorieva}}, \bibinfo {author}
  {\bibfnamefont {S.~V.}\ \bibnamefont {Dubonos}}, \ and\ \bibinfo {author}
  {\bibfnamefont {A.~A.}\ \bibnamefont {Firsov}},\ }\bibfield  {title}
  {\enquote {\bibinfo {title} {Two-dimensional gas of massless dirac fermions
  in graphene},}\ }\href@noop {} {\bibfield  {journal} {\bibinfo  {journal}
  {Nature}\ }\textbf {\bibinfo {volume} {438}},\ \bibinfo {pages} {197--200}
  (\bibinfo {year} {2005})}\BibitemShut {NoStop}%
\bibitem [{\citenamefont {Geim}\ and\ \citenamefont
  {Novoselov}(2007)}]{geim2007progress}%
  \BibitemOpen
  \bibfield  {author} {\bibinfo {author} {\bibfnamefont {A.}~\bibnamefont
  {Geim}}\ and\ \bibinfo {author} {\bibfnamefont {K.}~\bibnamefont
  {Novoselov}},\ }\bibfield  {title} {\enquote {\bibinfo {title} {The rise of
  graphene},}\ }\href@noop {} {\bibfield  {journal} {\bibinfo  {journal}
  {Nature Mat.}\ }\textbf {\bibinfo {volume} {6}},\ \bibinfo {pages} {183--191}
  (\bibinfo {year} {2007})}\BibitemShut {NoStop}%
\bibitem [{\citenamefont {Xu}\ \emph {et~al.}(2013)\citenamefont {Xu},
  \citenamefont {Liang}, \citenamefont {Shi},\ and\ \citenamefont
  {Chen}}]{xu2013_2Dmaterials}%
  \BibitemOpen
  \bibfield  {author} {\bibinfo {author} {\bibfnamefont {M.}~\bibnamefont
  {Xu}}, \bibinfo {author} {\bibfnamefont {T.}~\bibnamefont {Liang}}, \bibinfo
  {author} {\bibfnamefont {M.}~\bibnamefont {Shi}}, \ and\ \bibinfo {author}
  {\bibfnamefont {H.}~\bibnamefont {Chen}},\ }\bibfield  {title} {\enquote
  {\bibinfo {title} {Graphene-like two-dimensional materials},}\ }\href@noop {}
  {\bibfield  {journal} {\bibinfo  {journal} {Chemical reviews}\ }\textbf
  {\bibinfo {volume} {113}},\ \bibinfo {pages} {3766--3798} (\bibinfo {year}
  {2013})}\BibitemShut {NoStop}%
\bibitem [{\citenamefont {Dean}\ \emph {et~al.}(2010)\citenamefont {Dean},
  \citenamefont {Young}, \citenamefont {Meric}, \citenamefont {Lee},
  \citenamefont {Wang}, \citenamefont {Sorgenfrei}, \citenamefont {Watanabe},
  \citenamefont {Taniguchi}, \citenamefont {Kim}, \citenamefont {Shepard},\
  and\ \citenamefont {Hone}}]{dean2010hbntransfer}%
  \BibitemOpen
  \bibfield  {author} {\bibinfo {author} {\bibfnamefont {C.}~\bibnamefont
  {Dean}}, \bibinfo {author} {\bibfnamefont {A.}~\bibnamefont {Young}},
  \bibinfo {author} {\bibfnamefont {I.}~\bibnamefont {Meric}}, \bibinfo
  {author} {\bibfnamefont {C.}~\bibnamefont {Lee}}, \bibinfo {author}
  {\bibfnamefont {L.}~\bibnamefont {Wang}}, \bibinfo {author} {\bibfnamefont
  {S.}~\bibnamefont {Sorgenfrei}}, \bibinfo {author} {\bibfnamefont
  {K.}~\bibnamefont {Watanabe}}, \bibinfo {author} {\bibfnamefont
  {T.}~\bibnamefont {Taniguchi}}, \bibinfo {author} {\bibfnamefont
  {P.}~\bibnamefont {Kim}}, \bibinfo {author} {\bibfnamefont {K.}~\bibnamefont
  {Shepard}}, \ and\ \bibinfo {author} {\bibfnamefont {J.}~\bibnamefont
  {Hone}},\ }\bibfield  {title} {\enquote {\bibinfo {title} {Boron nitride
  substrates for high-quality graphene electronics},}\ }\href@noop {}
  {\bibfield  {journal} {\bibinfo  {journal} {Nat. Nanotech.}\ }\textbf
  {\bibinfo {volume} {5}},\ \bibinfo {pages} {722--726} (\bibinfo {year}
  {2010})}\BibitemShut {NoStop}%
\bibitem [{\citenamefont {Geim}\ and\ \citenamefont
  {Grigorieva}(2013)}]{geim2013van}%
  \BibitemOpen
  \bibfield  {author} {\bibinfo {author} {\bibfnamefont {A.~K.}\ \bibnamefont
  {Geim}}\ and\ \bibinfo {author} {\bibfnamefont {I.~V.}\ \bibnamefont
  {Grigorieva}},\ }\bibfield  {title} {\enquote {\bibinfo {title} {Van der
  waals heterostructures},}\ }\href@noop {} {\bibfield  {journal} {\bibinfo
  {journal} {Nature}\ }\textbf {\bibinfo {volume} {499}},\ \bibinfo {pages}
  {419--425} (\bibinfo {year} {2013})}\BibitemShut {NoStop}%
\bibitem [{\citenamefont {Novoselov}\ \emph {et~al.}(2016)\citenamefont
  {Novoselov}, \citenamefont {Mishchenko}, \citenamefont {Carvalho},\ and\
  \citenamefont {Neto}}]{novoselov2016_2d_vdW}%
  \BibitemOpen
  \bibfield  {author} {\bibinfo {author} {\bibfnamefont {K.}~\bibnamefont
  {Novoselov}}, \bibinfo {author} {\bibfnamefont {A.}~\bibnamefont
  {Mishchenko}}, \bibinfo {author} {\bibfnamefont {A.}~\bibnamefont
  {Carvalho}}, \ and\ \bibinfo {author} {\bibfnamefont {A.~C.}\ \bibnamefont
  {Neto}},\ }\bibfield  {title} {\enquote {\bibinfo {title} {2d materials and
  van der waals heterostructures},}\ }\href@noop {} {\bibfield  {journal}
  {\bibinfo  {journal} {Science}\ }\textbf {\bibinfo {volume} {353}},\ \bibinfo
  {pages} {aac9439} (\bibinfo {year} {2016})}\BibitemShut {NoStop}%
\bibitem [{\citenamefont {Baringhaus}\ \emph {et~al.}(2015)\citenamefont
  {Baringhaus}, \citenamefont {St{\"o}hr}, \citenamefont {Forti}, \citenamefont
  {Starke},\ and\ \citenamefont {Tegenkamp}}]{baringhaus2015ChemicalGating}%
  \BibitemOpen
  \bibfield  {author} {\bibinfo {author} {\bibfnamefont {J.}~\bibnamefont
  {Baringhaus}}, \bibinfo {author} {\bibfnamefont {A.}~\bibnamefont
  {St{\"o}hr}}, \bibinfo {author} {\bibfnamefont {S.}~\bibnamefont {Forti}},
  \bibinfo {author} {\bibfnamefont {U.}~\bibnamefont {Starke}}, \ and\ \bibinfo
  {author} {\bibfnamefont {C.}~\bibnamefont {Tegenkamp}},\ }\bibfield  {title}
  {\enquote {\bibinfo {title} {Ballistic bipolar junctions in chemically gated
  graphene ribbons},}\ }\href@noop {} {\bibfield  {journal} {\bibinfo
  {journal} {Scientific reports}\ }\textbf {\bibinfo {volume} {5}} (\bibinfo
  {year} {2015})}\BibitemShut {NoStop}%
\bibitem [{\citenamefont {Han}\ \emph {et~al.}(2007)\citenamefont {Han},
  \citenamefont {{\"O}zyilmaz}, \citenamefont {Zhang},\ and\ \citenamefont
  {Kim}}]{han2007gnr}%
  \BibitemOpen
  \bibfield  {author} {\bibinfo {author} {\bibfnamefont {M.~Y.}\ \bibnamefont
  {Han}}, \bibinfo {author} {\bibfnamefont {B.}~\bibnamefont {{\"O}zyilmaz}},
  \bibinfo {author} {\bibfnamefont {Y.}~\bibnamefont {Zhang}}, \ and\ \bibinfo
  {author} {\bibfnamefont {P.}~\bibnamefont {Kim}},\ }\bibfield  {title}
  {\enquote {\bibinfo {title} {Energy band-gap engineering of graphene
  nanoribbons},}\ }\href@noop {} {\bibfield  {journal} {\bibinfo  {journal}
  {Physical review letters}\ }\textbf {\bibinfo {volume} {98}},\ \bibinfo
  {pages} {206805} (\bibinfo {year} {2007})}\BibitemShut {NoStop}%
\bibitem [{\citenamefont {Tombros}\ \emph {et~al.}(2011)\citenamefont
  {Tombros}, \citenamefont {Veligura}, \citenamefont {Junesch}, \citenamefont
  {Guimaraes}, \citenamefont {Vera-Marun}, \citenamefont {Jonkman},\ and\
  \citenamefont {van Wees}}]{Tombros2011}%
  \BibitemOpen
  \bibfield  {author} {\bibinfo {author} {\bibfnamefont {N.}~\bibnamefont
  {Tombros}}, \bibinfo {author} {\bibfnamefont {A.}~\bibnamefont {Veligura}},
  \bibinfo {author} {\bibfnamefont {J.}~\bibnamefont {Junesch}}, \bibinfo
  {author} {\bibfnamefont {M.~H.~D.}\ \bibnamefont {Guimaraes}}, \bibinfo
  {author} {\bibfnamefont {I.~J.}\ \bibnamefont {Vera-Marun}}, \bibinfo
  {author} {\bibfnamefont {H.~T.}\ \bibnamefont {Jonkman}}, \ and\ \bibinfo
  {author} {\bibfnamefont {B.~J.}\ \bibnamefont {van Wees}},\ }\bibfield
  {title} {\enquote {\bibinfo {title} {Quantized conductance of a suspended
  graphene nanoconstriction},}\ }\href {http://dx.doi.org/10.1038/nphys2009}
  {\bibfield  {journal} {\bibinfo  {journal} {Nat Phys}\ }\textbf {\bibinfo
  {volume} {7}},\ \bibinfo {pages} {697--700} (\bibinfo {year}
  {2011})}\BibitemShut {NoStop}%
\bibitem [{\citenamefont {Sandner}\ \emph {et~al.}(2015)\citenamefont
  {Sandner}, \citenamefont {Preis}, \citenamefont {Schell}, \citenamefont
  {Giudici}, \citenamefont {Watanabe}, \citenamefont {Taniguchi}, \citenamefont
  {Weiss},\ and\ \citenamefont {Eroms}}]{sandner2015ballistic}%
  \BibitemOpen
  \bibfield  {author} {\bibinfo {author} {\bibfnamefont {A.}~\bibnamefont
  {Sandner}}, \bibinfo {author} {\bibfnamefont {T.}~\bibnamefont {Preis}},
  \bibinfo {author} {\bibfnamefont {C.}~\bibnamefont {Schell}}, \bibinfo
  {author} {\bibfnamefont {P.}~\bibnamefont {Giudici}}, \bibinfo {author}
  {\bibfnamefont {K.}~\bibnamefont {Watanabe}}, \bibinfo {author}
  {\bibfnamefont {T.}~\bibnamefont {Taniguchi}}, \bibinfo {author}
  {\bibfnamefont {D.}~\bibnamefont {Weiss}}, \ and\ \bibinfo {author}
  {\bibfnamefont {J.}~\bibnamefont {Eroms}},\ }\bibfield  {title} {\enquote
  {\bibinfo {title} {Ballistic transport in graphene antidot lattices},}\
  }\href@noop {} {\bibfield  {journal} {\bibinfo  {journal} {Nano letters}\
  }\textbf {\bibinfo {volume} {15}},\ \bibinfo {pages} {8402--8406} (\bibinfo
  {year} {2015})}\BibitemShut {NoStop}%
\bibitem [{\citenamefont {Yagi}\ \emph {et~al.}(2015)\citenamefont {Yagi},
  \citenamefont {Sakakibara}, \citenamefont {Ebisuoka}, \citenamefont {Onishi},
  \citenamefont {Watanabe}, \citenamefont {Taniguchi},\ and\ \citenamefont
  {Iye}}]{Yagi2015}%
  \BibitemOpen
  \bibfield  {author} {\bibinfo {author} {\bibfnamefont {R.}~\bibnamefont
  {Yagi}}, \bibinfo {author} {\bibfnamefont {R.}~\bibnamefont {Sakakibara}},
  \bibinfo {author} {\bibfnamefont {R.}~\bibnamefont {Ebisuoka}}, \bibinfo
  {author} {\bibfnamefont {J.}~\bibnamefont {Onishi}}, \bibinfo {author}
  {\bibfnamefont {K.}~\bibnamefont {Watanabe}}, \bibinfo {author}
  {\bibfnamefont {T.}~\bibnamefont {Taniguchi}}, \ and\ \bibinfo {author}
  {\bibfnamefont {Y.}~\bibnamefont {Iye}},\ }\bibfield  {title} {\enquote
  {\bibinfo {title} {Ballistic transport in graphene antidot lattices},}\
  }\href {http://link.aps.org/doi/10.1103/PhysRevB.92.195406} {\bibfield
  {journal} {\bibinfo  {journal} {Phys. Rev. B}\ }\textbf {\bibinfo {volume}
  {92}},\ \bibinfo {pages} {195406--} (\bibinfo {year} {2015})}\BibitemShut
  {NoStop}%
\bibitem [{\citenamefont {Klein}(1929)}]{klein1929reflexion}%
  \BibitemOpen
  \bibfield  {author} {\bibinfo {author} {\bibfnamefont {O.}~\bibnamefont
  {Klein}},\ }\bibfield  {title} {\enquote {\bibinfo {title} {Die reflexion von
  elektronen an einem potentialsprung nach der relativistischen dynamik von
  dirac},}\ }\href@noop {} {\bibfield  {journal} {\bibinfo  {journal} {Z.
  Phys.}\ }\textbf {\bibinfo {volume} {53}},\ \bibinfo {pages} {157--165}
  (\bibinfo {year} {1929})}\BibitemShut {NoStop}%
\bibitem [{\citenamefont {Katsnelson}, \citenamefont {Novoselov},\ and\
  \citenamefont {Geim}(2006)}]{katsnelson2006chiral}%
  \BibitemOpen
  \bibfield  {author} {\bibinfo {author} {\bibfnamefont {M.~I.}\ \bibnamefont
  {Katsnelson}}, \bibinfo {author} {\bibfnamefont {K.~S.}\ \bibnamefont
  {Novoselov}}, \ and\ \bibinfo {author} {\bibfnamefont {A.~K.}\ \bibnamefont
  {Geim}},\ }\bibfield  {title} {\enquote {\bibinfo {title} {Chiral tunnelling
  and the klein paradox in graphene},}\ }\href@noop {} {\bibfield  {journal}
  {\bibinfo  {journal} {Nat. Phys.}\ }\textbf {\bibinfo {volume} {2}},\
  \bibinfo {pages} {620--625} (\bibinfo {year} {2006})}\BibitemShut {NoStop}%
\bibitem [{\citenamefont {Cheianov}\ and\ \citenamefont
  {Fal'ko}(2006)}]{cheianov2006selective}%
  \BibitemOpen
  \bibfield  {author} {\bibinfo {author} {\bibfnamefont {V.~V.}\ \bibnamefont
  {Cheianov}}\ and\ \bibinfo {author} {\bibfnamefont {V.~I.}\ \bibnamefont
  {Fal'ko}},\ }\bibfield  {title} {\enquote {\bibinfo {title} {Selective
  transmission of dirac electrons and ballistic magnetoresistance of np
  junctions in graphene},}\ }\href@noop {} {\bibfield  {journal} {\bibinfo
  {journal} {Phys. Rev. B}\ }\textbf {\bibinfo {volume} {74}},\ \bibinfo
  {pages} {041403} (\bibinfo {year} {2006})}\BibitemShut {NoStop}%
\bibitem [{\citenamefont {Cheianov}, \citenamefont {Fal'ko},\ and\
  \citenamefont {Altshuler}(2007)}]{cheianov2007focusing}%
  \BibitemOpen
  \bibfield  {author} {\bibinfo {author} {\bibfnamefont {V.~V.}\ \bibnamefont
  {Cheianov}}, \bibinfo {author} {\bibfnamefont {V.}~\bibnamefont {Fal'ko}}, \
  and\ \bibinfo {author} {\bibfnamefont {B.}~\bibnamefont {Altshuler}},\
  }\bibfield  {title} {\enquote {\bibinfo {title} {The focusing of electron
  flow and a veselago lens in graphene pn junctions},}\ }\href@noop {}
  {\bibfield  {journal} {\bibinfo  {journal} {Science}\ }\textbf {\bibinfo
  {volume} {315}},\ \bibinfo {pages} {1252--1255} (\bibinfo {year}
  {2007})}\BibitemShut {NoStop}%
\bibitem [{\citenamefont {Chen}\ \emph {et~al.}(2016)\citenamefont {Chen},
  \citenamefont {Han}, \citenamefont {Elahi}, \citenamefont {Habib},
  \citenamefont {Wang}, \citenamefont {Wen}, \citenamefont {Gao}, \citenamefont
  {Taniguchi}, \citenamefont {Watanabe}, \citenamefont {Hone}, \citenamefont
  {Ghosh},\ and\ \citenamefont {Dean}}]{chen2016electronOptics}%
  \BibitemOpen
  \bibfield  {author} {\bibinfo {author} {\bibfnamefont {S.}~\bibnamefont
  {Chen}}, \bibinfo {author} {\bibfnamefont {Z.}~\bibnamefont {Han}}, \bibinfo
  {author} {\bibfnamefont {M.~M.}\ \bibnamefont {Elahi}}, \bibinfo {author}
  {\bibfnamefont {K.~M.}\ \bibnamefont {Habib}}, \bibinfo {author}
  {\bibfnamefont {L.}~\bibnamefont {Wang}}, \bibinfo {author} {\bibfnamefont
  {B.}~\bibnamefont {Wen}}, \bibinfo {author} {\bibfnamefont {Y.}~\bibnamefont
  {Gao}}, \bibinfo {author} {\bibfnamefont {T.}~\bibnamefont {Taniguchi}},
  \bibinfo {author} {\bibfnamefont {K.}~\bibnamefont {Watanabe}}, \bibinfo
  {author} {\bibfnamefont {J.}~\bibnamefont {Hone}}, \bibinfo {author}
  {\bibfnamefont {A.~W.}\ \bibnamefont {Ghosh}}, \ and\ \bibinfo {author}
  {\bibfnamefont {C.~R.}\ \bibnamefont {Dean}},\ }\bibfield  {title} {\enquote
  {\bibinfo {title} {Electron optics with pn junctions in ballistic
  graphene},}\ }\href@noop {} {\bibfield  {journal} {\bibinfo  {journal}
  {Science}\ }\textbf {\bibinfo {volume} {353}},\ \bibinfo {pages} {1522--1525}
  (\bibinfo {year} {2016})}\BibitemShut {NoStop}%
\bibitem [{\citenamefont {Liu}, \citenamefont {Gorini},\ and\ \citenamefont
  {Richter}(2017)}]{liu2017lensing}%
  \BibitemOpen
  \bibfield  {author} {\bibinfo {author} {\bibfnamefont {M.-H.}\ \bibnamefont
  {Liu}}, \bibinfo {author} {\bibfnamefont {C.}~\bibnamefont {Gorini}}, \ and\
  \bibinfo {author} {\bibfnamefont {K.}~\bibnamefont {Richter}},\ }\bibfield
  {title} {\enquote {\bibinfo {title} {Creating and steering highly directional
  electron beams in graphene},}\ }\href@noop {} {\bibfield  {journal} {\bibinfo
   {journal} {Physical Review Letters}\ }\textbf {\bibinfo {volume} {118}},\
  \bibinfo {pages} {066801} (\bibinfo {year} {2017})}\BibitemShut {NoStop}%
\bibitem [{\citenamefont {Buscema}\ \emph {et~al.}(2014)\citenamefont
  {Buscema}, \citenamefont {Groenendijk}, \citenamefont {Steele}, \citenamefont
  {van~der Zant},\ and\ \citenamefont
  {Castellanos-Gomez}}]{buscema2014photovoltaic}%
  \BibitemOpen
  \bibfield  {author} {\bibinfo {author} {\bibfnamefont {M.}~\bibnamefont
  {Buscema}}, \bibinfo {author} {\bibfnamefont {D.~J.}\ \bibnamefont
  {Groenendijk}}, \bibinfo {author} {\bibfnamefont {G.~A.}\ \bibnamefont
  {Steele}}, \bibinfo {author} {\bibfnamefont {H.~S.}\ \bibnamefont {van~der
  Zant}}, \ and\ \bibinfo {author} {\bibfnamefont {A.}~\bibnamefont
  {Castellanos-Gomez}},\ }\bibfield  {title} {\enquote {\bibinfo {title}
  {Photovoltaic effect in few-layer black phosphorus pn junctions defined by
  local electrostatic gating},}\ }\href@noop {} {\bibfield  {journal} {\bibinfo
   {journal} {Nature communications}\ }\textbf {\bibinfo {volume} {5}}
  (\bibinfo {year} {2014})}\BibitemShut {NoStop}%
\bibitem [{\citenamefont {Baugher}\ \emph {et~al.}(2014)\citenamefont
  {Baugher}, \citenamefont {Churchill}, \citenamefont {Yang},\ and\
  \citenamefont {Jarillo-Herrero}}]{baugher2014WSe2_pn-diode}%
  \BibitemOpen
  \bibfield  {author} {\bibinfo {author} {\bibfnamefont {B.~W.}\ \bibnamefont
  {Baugher}}, \bibinfo {author} {\bibfnamefont {H.~O.}\ \bibnamefont
  {Churchill}}, \bibinfo {author} {\bibfnamefont {Y.}~\bibnamefont {Yang}}, \
  and\ \bibinfo {author} {\bibfnamefont {P.}~\bibnamefont {Jarillo-Herrero}},\
  }\bibfield  {title} {\enquote {\bibinfo {title} {Optoelectronic devices based
  on electrically tunable pn diodes in a monolayer dichalcogenide},}\
  }\href@noop {} {\bibfield  {journal} {\bibinfo  {journal} {Nature
  nanotechnology}\ }\textbf {\bibinfo {volume} {9}},\ \bibinfo {pages}
  {262--267} (\bibinfo {year} {2014})}\BibitemShut {NoStop}%
\bibitem [{\citenamefont {Young}\ and\ \citenamefont
  {Kim}(2009)}]{young2009quantum}%
  \BibitemOpen
  \bibfield  {author} {\bibinfo {author} {\bibfnamefont {A.~F.}\ \bibnamefont
  {Young}}\ and\ \bibinfo {author} {\bibfnamefont {P.}~\bibnamefont {Kim}},\
  }\bibfield  {title} {\enquote {\bibinfo {title} {Quantum interference and
  klein tunnelling in graphene heterojunctions},}\ }\href@noop {} {\bibfield
  {journal} {\bibinfo  {journal} {Nat. Phys.}\ }\textbf {\bibinfo {volume}
  {5}},\ \bibinfo {pages} {222--226} (\bibinfo {year} {2009})}\BibitemShut
  {NoStop}%
\bibitem [{\citenamefont {Stander}, \citenamefont {Huard},\ and\ \citenamefont
  {Goldhaber-Gordon}(2009)}]{stander2009evidence}%
  \BibitemOpen
  \bibfield  {author} {\bibinfo {author} {\bibfnamefont {N.}~\bibnamefont
  {Stander}}, \bibinfo {author} {\bibfnamefont {B.}~\bibnamefont {Huard}}, \
  and\ \bibinfo {author} {\bibfnamefont {D.}~\bibnamefont {Goldhaber-Gordon}},\
  }\bibfield  {title} {\enquote {\bibinfo {title} {Evidence for klein tunneling
  in graphene pn junctions},}\ }\href@noop {} {\bibfield  {journal} {\bibinfo
  {journal} {Phys. Rev. Lett.}\ }\textbf {\bibinfo {volume} {102}},\ \bibinfo
  {pages} {026807} (\bibinfo {year} {2009})}\BibitemShut {NoStop}%
\bibitem [{\citenamefont {Williams}, \citenamefont {DiCarlo},\ and\
  \citenamefont {Marcus}(2007)}]{williams2007quantum}%
  \BibitemOpen
  \bibfield  {author} {\bibinfo {author} {\bibfnamefont {J.}~\bibnamefont
  {Williams}}, \bibinfo {author} {\bibfnamefont {L.}~\bibnamefont {DiCarlo}}, \
  and\ \bibinfo {author} {\bibfnamefont {C.}~\bibnamefont {Marcus}},\
  }\bibfield  {title} {\enquote {\bibinfo {title} {Quantum hall effect in a
  gate-controlled pn junction of graphene},}\ }\href@noop {} {\bibfield
  {journal} {\bibinfo  {journal} {Science}\ }\textbf {\bibinfo {volume}
  {317}},\ \bibinfo {pages} {638--641} (\bibinfo {year} {2007})}\BibitemShut
  {NoStop}%
\bibitem [{\citenamefont {\"Ozyilmaz}\ \emph {et~al.}(2007)\citenamefont
  {\"Ozyilmaz}, \citenamefont {Jarillo-Herrero}, \citenamefont {Efetov},
  \citenamefont {Abanin}, \citenamefont {Levitov},\ and\ \citenamefont
  {Kim}}]{ozyilmaz2007electronic}%
  \BibitemOpen
  \bibfield  {author} {\bibinfo {author} {\bibfnamefont {B.}~\bibnamefont
  {\"Ozyilmaz}}, \bibinfo {author} {\bibfnamefont {P.}~\bibnamefont
  {Jarillo-Herrero}}, \bibinfo {author} {\bibfnamefont {D.}~\bibnamefont
  {Efetov}}, \bibinfo {author} {\bibfnamefont {D.~A.}\ \bibnamefont {Abanin}},
  \bibinfo {author} {\bibfnamefont {L.~S.}\ \bibnamefont {Levitov}}, \ and\
  \bibinfo {author} {\bibfnamefont {P.}~\bibnamefont {Kim}},\ }\bibfield
  {title} {\enquote {\bibinfo {title} {Electronic transport and quantum hall
  effect in bipolar graphene pnp junctions},}\ }\href@noop {} {\bibfield
  {journal} {\bibinfo  {journal} {Phys. Rev. Lett.}\ }\textbf {\bibinfo
  {volume} {99}},\ \bibinfo {pages} {166804} (\bibinfo {year}
  {2007})}\BibitemShut {NoStop}%
\bibitem [{\citenamefont {Drienovsky}\ \emph {et~al.}(2014)\citenamefont
  {Drienovsky}, \citenamefont {Schrettenbrunner}, \citenamefont {Sandner},
  \citenamefont {Weiss}, \citenamefont {Eroms}, \citenamefont {Liu},
  \citenamefont {Tkatschenko},\ and\ \citenamefont
  {Richter}}]{drienovsky2014multibarriers}%
  \BibitemOpen
  \bibfield  {author} {\bibinfo {author} {\bibfnamefont {M.}~\bibnamefont
  {Drienovsky}}, \bibinfo {author} {\bibfnamefont {F.-X.}\ \bibnamefont
  {Schrettenbrunner}}, \bibinfo {author} {\bibfnamefont {A.}~\bibnamefont
  {Sandner}}, \bibinfo {author} {\bibfnamefont {D.}~\bibnamefont {Weiss}},
  \bibinfo {author} {\bibfnamefont {J.}~\bibnamefont {Eroms}}, \bibinfo
  {author} {\bibfnamefont {M.-H.}\ \bibnamefont {Liu}}, \bibinfo {author}
  {\bibfnamefont {F.}~\bibnamefont {Tkatschenko}}, \ and\ \bibinfo {author}
  {\bibfnamefont {K.}~\bibnamefont {Richter}},\ }\bibfield  {title} {\enquote
  {\bibinfo {title} {Towards superlattices: Lateral bipolar multibarriers in
  graphene},}\ }\href {\doibase 10.1103/PhysRevB.89.115421} {\bibfield
  {journal} {\bibinfo  {journal} {Phys. Rev. B}\ }\textbf {\bibinfo {volume}
  {89}},\ \bibinfo {pages} {115421} (\bibinfo {year} {2014})}\BibitemShut
  {NoStop}%
\bibitem [{\citenamefont {Fallahazad}\ \emph {et~al.}(2010)\citenamefont
  {Fallahazad}, \citenamefont {Kim}, \citenamefont {Colombo},\ and\
  \citenamefont {Tutuc}}]{fallahazad2010impurities}%
  \BibitemOpen
  \bibfield  {author} {\bibinfo {author} {\bibfnamefont {B.}~\bibnamefont
  {Fallahazad}}, \bibinfo {author} {\bibfnamefont {S.}~\bibnamefont {Kim}},
  \bibinfo {author} {\bibfnamefont {L.}~\bibnamefont {Colombo}}, \ and\
  \bibinfo {author} {\bibfnamefont {E.}~\bibnamefont {Tutuc}},\ }\bibfield
  {title} {\enquote {\bibinfo {title} {Dielectric thickness dependence of
  carrier mobility in graphene with hfo2 top dielectric},}\ }\href@noop {}
  {\bibfield  {journal} {\bibinfo  {journal} {Applied Physics Letters}\
  }\textbf {\bibinfo {volume} {97}},\ \bibinfo {pages} {123105} (\bibinfo
  {year} {2010})}\BibitemShut {NoStop}%
\bibitem [{\citenamefont {Wang}\ \emph {et~al.}(2013)\citenamefont {Wang},
  \citenamefont {Meric}, \citenamefont {Huang}, \citenamefont {Gao},
  \citenamefont {Gao}, \citenamefont {Tran}, \citenamefont {Taniguchi},
  \citenamefont {Watanabe}, \citenamefont {Campos}, \citenamefont {Muller},
  \citenamefont {Guo}, \citenamefont {Kim}, \citenamefont {Hone}, \citenamefont
  {Shepard},\ and\ \citenamefont {Dean}}]{wang2013oneDcontacts}%
  \BibitemOpen
  \bibfield  {author} {\bibinfo {author} {\bibfnamefont {L.}~\bibnamefont
  {Wang}}, \bibinfo {author} {\bibfnamefont {I.}~\bibnamefont {Meric}},
  \bibinfo {author} {\bibfnamefont {P.}~\bibnamefont {Huang}}, \bibinfo
  {author} {\bibfnamefont {Q.}~\bibnamefont {Gao}}, \bibinfo {author}
  {\bibfnamefont {Y.}~\bibnamefont {Gao}}, \bibinfo {author} {\bibfnamefont
  {H.}~\bibnamefont {Tran}}, \bibinfo {author} {\bibfnamefont {T.}~\bibnamefont
  {Taniguchi}}, \bibinfo {author} {\bibfnamefont {K.}~\bibnamefont {Watanabe}},
  \bibinfo {author} {\bibfnamefont {L.}~\bibnamefont {Campos}}, \bibinfo
  {author} {\bibfnamefont {D.}~\bibnamefont {Muller}}, \bibinfo {author}
  {\bibfnamefont {J.}~\bibnamefont {Guo}}, \bibinfo {author} {\bibfnamefont
  {P.}~\bibnamefont {Kim}}, \bibinfo {author} {\bibfnamefont {J.}~\bibnamefont
  {Hone}}, \bibinfo {author} {\bibfnamefont {K.}~\bibnamefont {Shepard}}, \
  and\ \bibinfo {author} {\bibfnamefont {C.}~\bibnamefont {Dean}},\ }\bibfield
  {title} {\enquote {\bibinfo {title} {One-dimensional electrical contact to a
  two-dimensional material},}\ }\href@noop {} {\bibfield  {journal} {\bibinfo
  {journal} {Science}\ }\textbf {\bibinfo {volume} {342}},\ \bibinfo {pages}
  {614--617} (\bibinfo {year} {2013})}\BibitemShut {NoStop}%
\bibitem [{\citenamefont {Ponomarenko}\ \emph {et~al.}(2011)\citenamefont
  {Ponomarenko}, \citenamefont {Geim}, \citenamefont {Zhukov}, \citenamefont
  {Jalil}, \citenamefont {Morozov}, \citenamefont {Novoselov}, \citenamefont
  {Grigorieva}, \citenamefont {Hill}, \citenamefont {Cheianov}, \citenamefont
  {Fal'Ko}, \citenamefont {Watanabe}, \citenamefont {Taniguchi},\ and\
  \citenamefont {Gorbachev}}]{ponomarenko2011DoubleLayers}%
  \BibitemOpen
  \bibfield  {author} {\bibinfo {author} {\bibfnamefont {L.}~\bibnamefont
  {Ponomarenko}}, \bibinfo {author} {\bibfnamefont {A.}~\bibnamefont {Geim}},
  \bibinfo {author} {\bibfnamefont {A.}~\bibnamefont {Zhukov}}, \bibinfo
  {author} {\bibfnamefont {R.}~\bibnamefont {Jalil}}, \bibinfo {author}
  {\bibfnamefont {S.}~\bibnamefont {Morozov}}, \bibinfo {author} {\bibfnamefont
  {K.}~\bibnamefont {Novoselov}}, \bibinfo {author} {\bibfnamefont
  {I.}~\bibnamefont {Grigorieva}}, \bibinfo {author} {\bibfnamefont
  {E.}~\bibnamefont {Hill}}, \bibinfo {author} {\bibfnamefont {V.}~\bibnamefont
  {Cheianov}}, \bibinfo {author} {\bibfnamefont {V.}~\bibnamefont {Fal'Ko}},
  \bibinfo {author} {\bibfnamefont {K.}~\bibnamefont {Watanabe}}, \bibinfo
  {author} {\bibfnamefont {T.}~\bibnamefont {Taniguchi}}, \ and\ \bibinfo
  {author} {\bibfnamefont {R.}~\bibnamefont {Gorbachev}},\ }\bibfield  {title}
  {\enquote {\bibinfo {title} {Tunable metal-insulator transition in
  double-layer graphene heterostructures},}\ }\href@noop {} {\bibfield
  {journal} {\bibinfo  {journal} {Nature Physics}\ }\textbf {\bibinfo {volume}
  {7}},\ \bibinfo {pages} {958--961} (\bibinfo {year} {2011})}\BibitemShut
  {NoStop}%
\bibitem [{\citenamefont {Hunt}\ \emph {et~al.}(2013)\citenamefont {Hunt},
  \citenamefont {Sanchez-Yamagishi}, \citenamefont {Young}, \citenamefont
  {Yankowitz}, \citenamefont {LeRoy}, \citenamefont {Watanabe}, \citenamefont
  {Taniguchi}, \citenamefont {Moon}, \citenamefont {Koshino}, \citenamefont
  {Jarillo-Herrero},\ and\ \citenamefont
  {Ashoori}}]{hunt2013massiveDF_Hofstadter}%
  \BibitemOpen
  \bibfield  {author} {\bibinfo {author} {\bibfnamefont {B.}~\bibnamefont
  {Hunt}}, \bibinfo {author} {\bibfnamefont {J.}~\bibnamefont
  {Sanchez-Yamagishi}}, \bibinfo {author} {\bibfnamefont {A.}~\bibnamefont
  {Young}}, \bibinfo {author} {\bibfnamefont {M.}~\bibnamefont {Yankowitz}},
  \bibinfo {author} {\bibfnamefont {B.~J.}\ \bibnamefont {LeRoy}}, \bibinfo
  {author} {\bibfnamefont {K.}~\bibnamefont {Watanabe}}, \bibinfo {author}
  {\bibfnamefont {T.}~\bibnamefont {Taniguchi}}, \bibinfo {author}
  {\bibfnamefont {P.}~\bibnamefont {Moon}}, \bibinfo {author} {\bibfnamefont
  {M.}~\bibnamefont {Koshino}}, \bibinfo {author} {\bibfnamefont
  {P.}~\bibnamefont {Jarillo-Herrero}}, \ and\ \bibinfo {author} {\bibfnamefont
  {R.}~\bibnamefont {Ashoori}},\ }\bibfield  {title} {\enquote {\bibinfo
  {title} {Massive dirac fermions and hofstadter butterfly in a van der waals
  heterostructure},}\ }\href@noop {} {\bibfield  {journal} {\bibinfo  {journal}
  {Science}\ }\textbf {\bibinfo {volume} {340}},\ \bibinfo {pages} {1427--1430}
  (\bibinfo {year} {2013})}\BibitemShut {NoStop}%
\bibitem [{\citenamefont {Nam}\ \emph {et~al.}(2011)\citenamefont {Nam},
  \citenamefont {Ki}, \citenamefont {Park}, \citenamefont {Kim}, \citenamefont
  {Kim},\ and\ \citenamefont {Lee}}]{nam2011embedded_gates}%
  \BibitemOpen
  \bibfield  {author} {\bibinfo {author} {\bibfnamefont {S.-G.}\ \bibnamefont
  {Nam}}, \bibinfo {author} {\bibfnamefont {D.-K.}\ \bibnamefont {Ki}},
  \bibinfo {author} {\bibfnamefont {J.~W.}\ \bibnamefont {Park}}, \bibinfo
  {author} {\bibfnamefont {Y.}~\bibnamefont {Kim}}, \bibinfo {author}
  {\bibfnamefont {J.~S.}\ \bibnamefont {Kim}}, \ and\ \bibinfo {author}
  {\bibfnamefont {H.-J.}\ \bibnamefont {Lee}},\ }\bibfield  {title} {\enquote
  {\bibinfo {title} {Ballistic transport of graphene pnp junctions with
  embedded local gates},}\ }\href@noop {} {\bibfield  {journal} {\bibinfo
  {journal} {Nanotechnology}\ }\textbf {\bibinfo {volume} {22}},\ \bibinfo
  {pages} {415203} (\bibinfo {year} {2011})}\BibitemShut {NoStop}%
\bibitem [{\citenamefont {Pereira}\ and\ \citenamefont
  {Neto}(2009)}]{pereira2009strain}%
  \BibitemOpen
  \bibfield  {author} {\bibinfo {author} {\bibfnamefont {V.~M.}\ \bibnamefont
  {Pereira}}\ and\ \bibinfo {author} {\bibfnamefont {A.~C.}\ \bibnamefont
  {Neto}},\ }\bibfield  {title} {\enquote {\bibinfo {title} {Strain engineering
  of graphene's electronic structure},}\ }\href@noop {} {\bibfield  {journal}
  {\bibinfo  {journal} {Physical Review Letters}\ }\textbf {\bibinfo {volume}
  {103}},\ \bibinfo {pages} {046801} (\bibinfo {year} {2009})}\BibitemShut
  {NoStop}%
\bibitem [{\citenamefont {Rickhaus}\ \emph {et~al.}(2013)\citenamefont
  {Rickhaus}, \citenamefont {Maurand}, \citenamefont {Liu}, \citenamefont
  {Weiss}, \citenamefont {Richter},\ and\ \citenamefont
  {Sch\"onenberger}}]{rickhaus2013ballistic}%
  \BibitemOpen
  \bibfield  {author} {\bibinfo {author} {\bibfnamefont {P.}~\bibnamefont
  {Rickhaus}}, \bibinfo {author} {\bibfnamefont {R.}~\bibnamefont {Maurand}},
  \bibinfo {author} {\bibfnamefont {M.-H.}\ \bibnamefont {Liu}}, \bibinfo
  {author} {\bibfnamefont {M.}~\bibnamefont {Weiss}}, \bibinfo {author}
  {\bibfnamefont {K.}~\bibnamefont {Richter}}, \ and\ \bibinfo {author}
  {\bibfnamefont {C.}~\bibnamefont {Sch\"onenberger}},\ }\bibfield  {title}
  {\enquote {\bibinfo {title} {Ballistic interferences in suspended
  graphene},}\ }\href@noop {} {\bibfield  {journal} {\bibinfo  {journal} {Nat.
  Commun.}\ }\textbf {\bibinfo {volume} {4}},\ \bibinfo {pages} {2342}
  (\bibinfo {year} {2013})}\BibitemShut {NoStop}%
\bibitem [{\citenamefont {Grushina}, \citenamefont {Ki},\ and\ \citenamefont
  {Morpurgo}(2013)}]{grushina2013ballistic}%
  \BibitemOpen
  \bibfield  {author} {\bibinfo {author} {\bibfnamefont {A.~L.}\ \bibnamefont
  {Grushina}}, \bibinfo {author} {\bibfnamefont {D.-K.}\ \bibnamefont {Ki}}, \
  and\ \bibinfo {author} {\bibfnamefont {A.~F.}\ \bibnamefont {Morpurgo}},\
  }\bibfield  {title} {\enquote {\bibinfo {title} {A ballistic pn junction in
  suspended graphene with split bottom gates},}\ }\href@noop {} {\bibfield
  {journal} {\bibinfo  {journal} {Appl. Phys. Lett.}\ }\textbf {\bibinfo
  {volume} {102}},\ \bibinfo {pages} {223102} (\bibinfo {year}
  {2013})}\BibitemShut {NoStop}%
\bibitem [{\citenamefont {Oksanen}\ \emph {et~al.}(2013)\citenamefont
  {Oksanen}, \citenamefont {Uppstu}, \citenamefont {Laitinen}, \citenamefont
  {Cox}, \citenamefont {Craciun}, \citenamefont {Russo}, \citenamefont
  {Harju},\ and\ \citenamefont {Hakonen}}]{oksanen2013single}%
  \BibitemOpen
  \bibfield  {author} {\bibinfo {author} {\bibfnamefont {M.}~\bibnamefont
  {Oksanen}}, \bibinfo {author} {\bibfnamefont {A.}~\bibnamefont {Uppstu}},
  \bibinfo {author} {\bibfnamefont {A.}~\bibnamefont {Laitinen}}, \bibinfo
  {author} {\bibfnamefont {D.~J.}\ \bibnamefont {Cox}}, \bibinfo {author}
  {\bibfnamefont {M.}~\bibnamefont {Craciun}}, \bibinfo {author} {\bibfnamefont
  {S.}~\bibnamefont {Russo}}, \bibinfo {author} {\bibfnamefont
  {A.}~\bibnamefont {Harju}}, \ and\ \bibinfo {author} {\bibfnamefont
  {P.}~\bibnamefont {Hakonen}},\ }\bibfield  {title} {\enquote {\bibinfo
  {title} {Single-and multi-mode fabry-p$\backslash$'erot interference in
  suspended graphene},}\ }\href@noop {} {\bibfield  {journal} {\bibinfo
  {journal} {arXiv: 1306.1212}\ } (\bibinfo {year} {2013})}\BibitemShut
  {NoStop}%
\bibitem [{\citenamefont {Rickhaus}\ \emph {et~al.}(2015)\citenamefont
  {Rickhaus}, \citenamefont {Liu}, \citenamefont {Makk}, \citenamefont
  {Maurand}, \citenamefont {Hess}, \citenamefont {Zihlmann}, \citenamefont
  {Weiss}, \citenamefont {Richter},\ and\ \citenamefont
  {Sch{\"o}nenberger}}]{rickhaus2015guiding}%
  \BibitemOpen
  \bibfield  {author} {\bibinfo {author} {\bibfnamefont {P.}~\bibnamefont
  {Rickhaus}}, \bibinfo {author} {\bibfnamefont {M.-H.}\ \bibnamefont {Liu}},
  \bibinfo {author} {\bibfnamefont {P.}~\bibnamefont {Makk}}, \bibinfo {author}
  {\bibfnamefont {R.}~\bibnamefont {Maurand}}, \bibinfo {author} {\bibfnamefont
  {S.}~\bibnamefont {Hess}}, \bibinfo {author} {\bibfnamefont {S.}~\bibnamefont
  {Zihlmann}}, \bibinfo {author} {\bibfnamefont {M.}~\bibnamefont {Weiss}},
  \bibinfo {author} {\bibfnamefont {K.}~\bibnamefont {Richter}}, \ and\
  \bibinfo {author} {\bibfnamefont {C.}~\bibnamefont {Sch{\"o}nenberger}},\
  }\bibfield  {title} {\enquote {\bibinfo {title} {Guiding of electrons in a
  few-mode ballistic graphene channel},}\ }\href@noop {} {\bibfield  {journal}
  {\bibinfo  {journal} {Nano letters}\ }\textbf {\bibinfo {volume} {15}},\
  \bibinfo {pages} {5819--5825} (\bibinfo {year} {2015})}\BibitemShut {NoStop}%
\bibitem [{Note1()}]{Note1}%
  \BibitemOpen
  \bibinfo {note} {Which would be six times the interlayer constant of
  graphene. This discrepancy can either be attributed to an overestimation of
  the height by the atomic force microscope (AFM) or a physical etching of the
  SiO$_2$ surface by O$_2$-plasma at higher powers and durations}\BibitemShut
  {NoStop}%
\bibitem [{\citenamefont {Dubey}\ \emph {et~al.}(2013)\citenamefont {Dubey},
  \citenamefont {Singh}, \citenamefont {Bhat}, \citenamefont {Parikh},
  \citenamefont {Grover}, \citenamefont {Sensarma}, \citenamefont {Tripathi},
  \citenamefont {Sengupta},\ and\ \citenamefont {Deshmukh}}]{dubey2013tunable}%
  \BibitemOpen
  \bibfield  {author} {\bibinfo {author} {\bibfnamefont {S.}~\bibnamefont
  {Dubey}}, \bibinfo {author} {\bibfnamefont {V.}~\bibnamefont {Singh}},
  \bibinfo {author} {\bibfnamefont {A.~K.}\ \bibnamefont {Bhat}}, \bibinfo
  {author} {\bibfnamefont {P.}~\bibnamefont {Parikh}}, \bibinfo {author}
  {\bibfnamefont {S.}~\bibnamefont {Grover}}, \bibinfo {author} {\bibfnamefont
  {R.}~\bibnamefont {Sensarma}}, \bibinfo {author} {\bibfnamefont
  {V.}~\bibnamefont {Tripathi}}, \bibinfo {author} {\bibfnamefont
  {K.}~\bibnamefont {Sengupta}}, \ and\ \bibinfo {author} {\bibfnamefont
  {M.~M.}\ \bibnamefont {Deshmukh}},\ }\bibfield  {title} {\enquote {\bibinfo
  {title} {Tunable superlattice in graphene to control the number of dirac
  points},}\ }\href@noop {} {\bibfield  {journal} {\bibinfo  {journal} {Nano
  Lett.}\ }\textbf {\bibinfo {volume} {13}},\ \bibinfo {pages} {3990--3995}
  (\bibinfo {year} {2013})}\BibitemShut {NoStop}%
\bibitem [{\citenamefont {Liu}\ and\ \citenamefont
  {Richter}(2012)}]{liu2012quantum_simulation}%
  \BibitemOpen
  \bibfield  {author} {\bibinfo {author} {\bibfnamefont {M.-H.}\ \bibnamefont
  {Liu}}\ and\ \bibinfo {author} {\bibfnamefont {K.}~\bibnamefont {Richter}},\
  }\bibfield  {title} {\enquote {\bibinfo {title} {Efficient quantum transport
  simulation for bulk graphene heterojunctions},}\ }\href {\doibase
  10.1103/PhysRevB.86.115455} {\bibfield  {journal} {\bibinfo  {journal} {Phys.
  Rev. B}\ }\textbf {\bibinfo {volume} {86}},\ \bibinfo {pages} {115455}
  (\bibinfo {year} {2012})}\BibitemShut {NoStop}%
\bibitem [{\citenamefont {Liu}\ \emph {et~al.}(2015)\citenamefont {Liu},
  \citenamefont {Rickhaus}, \citenamefont {Makk}, \citenamefont {T\'ov\'ari},
  \citenamefont {Maurand}, \citenamefont {Tkatschenko}, \citenamefont {Weiss},
  \citenamefont {Sch\"onenberger},\ and\ \citenamefont
  {Richter}}]{liu2015tight_binding}%
  \BibitemOpen
  \bibfield  {author} {\bibinfo {author} {\bibfnamefont {M.-H.}\ \bibnamefont
  {Liu}}, \bibinfo {author} {\bibfnamefont {P.}~\bibnamefont {Rickhaus}},
  \bibinfo {author} {\bibfnamefont {P.}~\bibnamefont {Makk}}, \bibinfo {author}
  {\bibfnamefont {E.}~\bibnamefont {T\'ov\'ari}}, \bibinfo {author}
  {\bibfnamefont {R.}~\bibnamefont {Maurand}}, \bibinfo {author} {\bibfnamefont
  {F.}~\bibnamefont {Tkatschenko}}, \bibinfo {author} {\bibfnamefont
  {M.}~\bibnamefont {Weiss}}, \bibinfo {author} {\bibfnamefont
  {C.}~\bibnamefont {Sch\"onenberger}}, \ and\ \bibinfo {author} {\bibfnamefont
  {K.}~\bibnamefont {Richter}},\ }\bibfield  {title} {\enquote {\bibinfo
  {title} {Scalable tight-binding model for graphene},}\ }\href {\doibase
  10.1103/PhysRevLett.114.036601} {\bibfield  {journal} {\bibinfo  {journal}
  {Phys. Rev. Lett.}\ }\textbf {\bibinfo {volume} {114}},\ \bibinfo {pages}
  {036601} (\bibinfo {year} {2015})}\BibitemShut {NoStop}%
\end{thebibliography}%

\end{document}